%% file: ms.tex
\documentclass[conference]{IEEEtran}
\IEEEoverridecommandlockouts
\usepackage{amsmath,amssymb,amsfonts}
\usepackage{algorithmic}
\usepackage{graphicx}
\usepackage{textcomp}
\usepackage{xcolor}
\usepackage[
    backend=biber,
    natbib=true,
    url=true, 
    doi=false,
    eprint=false,
    maxcitenames=99,
    maxnames=99,
    giveninits=true
]{biblatex}

\usepackage{booktabs}
\usepackage{xcolor}

\usepackage{xspace}

\usepackage{amsmath} 
\usepackage{amsthm}
\usepackage{amssymb}

\usepackage[disable]{todonotes}

\usepackage[mode=buildnew]{standalone}
\usepackage{pgfplots}

\usepackage{listings}
\usepackage{makecell}
\usepackage{nicefrac}


\def\BibTeX{{\rm B\kern-.05em{\sc i\kern-.025em b}\kern-.08em
    T\kern-.1667em\lower.7ex\hbox{E}\kern-.125emX}}
\begin{document}

\title{A Framework for Blockchain Interoperability and Runtime Selection} 

\author{\IEEEauthorblockN{Philipp Frauenthaler\IEEEauthorrefmark{1}, Michael Borkowski\IEEEauthorrefmark{1}, Stefan Schulte\IEEEauthorrefmark{1}}
\IEEEauthorblockA{\IEEEauthorrefmark{1} \textit{Distributed Systems Group, TU Wien, Vienna, Austria} \\
\{p.frauenthaler, m.borkowski, s.schulte\}@infosys.tuwien.ac.at}
}

\maketitle

\begin{abstract}
\input{sections/00_abstract.tex}
\end{abstract}


\input{sections/01_intro.tex}
\input{sections/02_background.tex}
\input{sections/03_approach.tex}
\input{sections/04_eval.tex}
\input{sections/06_related.tex}
\input{sections/07_conclusion.tex}



\printbibliography

\end{document}

%% file: sections/00_abstract.tex
The suitability of a particular blockchain for a given use case depends mainly on the blockchain's functional and non-functional properties. Such properties may vary over time, and thus, a selected blockchain may become unsuitable for a given use case. This uncertainty may hinder the widespread adoption of blockchain technologies in general. 

To mitigate the impact of volatile blockchain properties, we propose a framework that monitors several blockchains, allows the user to define functional and non-functional requirements, determines the most appropriate blockchain, and enables the switchover to that chain at runtime. Our evaluation using a reference implementation shows that switching to another blockchain can save cost and enable users to benefit from better performance and a higher level of trust. 

%% file: sections/01_intro.tex
\section{Introduction}
\label{sec:intro}
In the past few years, cryptocurrencies have gained significant public attention~ \cite{introduction-to-cryptocurrencies,bitcoin-under-the-hood}. The first and most prominent cryptocurrency is Bitcoin, proposed in 2008 by Satoshi Nakamoto~\cite{sok-challenges-for-bitcoin,bitcoin-whitepaper,bitcoin-survey}. While blockchains have proven to be suitable as distributed ledgers for recording transactions in Bitcoin and other cryptocurrencies~\cite{bitcoin-and-crpytocurrencies-technologies, bitcoin-survey}, blockchain technologies have also the potential to be applied in other use cases, e.g., the Internet of Things or business processes~\cite{christidis16, MWA+18, runtime-verification-business-process}.

The suitability of a particular blockchain for a given use case depends on various properties, e.g., the cost of writing data into that blockchain, the time until a data record is permanently included and thus remains unchanged with sufficient probability, the transaction throughput, the network's overall hash rate, or the distribution of the hash power among miners and mining pools~\cite{xu2017taxonomy}. Blockchain properties vary over time, e.g., the network hash rate may decrease\footnote{\url{https://etherscan.io/chart/hashrate}}. Variations of properties may cause a blockchain to become unsuitable for a given use case and, in further consequence, may hinder the widespread adoption of blockchain technologies in general, since the uncertain suitability for a given use case in the future poses significant risk for engineers evaluating the utilization of blockchains~\cite{blockchain-governance}.

To facilitate the adoption of blockchain technologies, we introduce a general-purpose framework for storing arbitrary data on blockchains. The framework abstracts technical details and offers interfaces for reading data from and writing data into multiple blockchains. To mitigate the impact of volatile blockchain properties, the proposed framework provides a switchover functionality allowing to switch to another, more beneficial blockchain at runtime. The framework monitors multiple blockchains, calculates their individual benefits and determines the most beneficial one based on user-defined requirements. Furthermore, the framework is able to react to various events such as a rapid decrease of a blockchain network's hash rate or a steadily increase of the cost of writing data into a blockchain. Beyond volatile blockchain properties, the proposed framework is also able to meet changing user demands by selecting a more appropriate blockchain. The combination of the blockchain selection algorithm and the switchover functionality enables users to benefit from low cost, better performance, and a higher level of trust. We use a reference implementation supporting Bitcoin, Ethereum, Ethereum Classic, and Expanse to evaluate our framework.

Summarizing, the contributions of our work are as follows:
\begin{itemize}
	\item We identify concrete metrics relevant for the monitoring and the runtime selection of blockchains.
	\item We propose a mechanism for determining the most beneficial blockchain based on user-defined requirements.
	\item We describe the requirements and the technical design of the proposed framework.
	\item We evaluate the benefits of the proposed framework in terms of cost, performance, and trust using a reference implementation.
\end{itemize}

The remainder of this paper is structured as follows: In Section~\ref{sec:background}, we further motivate our work. Section~\ref{sec:approach} presents relevant blockchain metrics, a mechanism for determining the most beneficial blockchain, and the switchover functionality. Section~\ref{sec:eval} provides an evaluation of the presented work and Section~\ref{sec:related} gives an overview of related work. Finally, Section~\ref{sec:conclusion} concludes the paper.

%% file: sections/02_background.tex
\section{Motivation}
\label{sec:background}

Our work aims to overcome issues regarding volatile blockchain properties, changing user demands, and the selection of an appropriate blockchain among many. In the following, we elaborate on these issues in more detail.

In the recent past, cryptocurrency users witnessed several price fluctuations, e.g., Bitcoin's market price reached an all-time high close to 20,000 USD in December 2017~\cite{bitcoin-basic-problem} and declined about 75\% from this peak until November 2018~\cite{bitcoin-75-decline}. The sensitivity of cryptocurrencies for price fluctuations influences the cost of writing data into a blockchain, e.g., through varying transaction fees.

Furthermore, in November 2018, the community has witnessed a ``hash war'' between the supporters of two competing hard forks of Bitcoin Cash (Bitcoin Cash ABC and Bitcoin Cash SV)~\cite{btch-hash-war}. To prevent the own fork to get damaged or even destroyed by the competitors, both opposing sides collected as much hash power as possible, leading to a centralization of both hard forks \cite{btch-hash-war-centralization}. Additionally, a considerable amount of hash power has been shifted from Bitcoin to Bitcoin Cash during the peak time of the ``war''. Due to this shift, Bitcoin's hash rate decreased by seven percent~\cite{btch-hash-war}. In general, a decreasing network hash rate may lead to a loss of trust in a blockchain since it becomes easier for malicious nodes to get control over more than 50\% of the overall hash power, enabling them to perform a 51\%~attack~\cite{blockchain-security-challenges}. Since changes leading to forks are constantly encouraged by community members, such conflicts may also emerge in the future. 

Another important aspect of a blockchain is its degree of decentralization. Many miners collude with others through mining pools \cite{bitcoin-and-crpytocurrencies-technologies}. If the number of participating nodes grows over time, the pool's overall hash rate increases as well. Mining pools concentrating more than 50\% of the hash power can perform any strategy available to a single majority miner~\cite{sok-challenges-for-bitcoin}. 

In blockchain networks, all transactions are replicated on every network node, increasing storage requirements and thus affecting scalability~\cite{xu2017taxonomy}.
Currently, on average, public blockchains like Bitcoin or Ethereum can only process 3--20 transactions per second~\cite{blockchain-software-connector}. Despite multiple attempts to solve this problem (e.g., increasing the size limit of Bitcoin blocks or transferring values off-chain by using the Lightning Network\footnote{\url{https://lightning.network/}}), scalability is still an open issue~\cite{blockchain-scalability-challenges, blockchain-software-connector}. Thus, an increasing workload may raise the time it takes until new transactions are mined. Additionally, users may compete more intensively to get their transactions included in one of the next blocks, possibly leading to higher transaction fees. Since it is difficult to predict the workload a blockchain is confronted with, the progression of cost and performance is unclear.

A blockchain project aiming to improve scalability is Corda, developed by the R3 consortium\footnote{\url{https://www.r3.com/}}. The consortium addresses the scalability problem by reducing the replication of transactions across network nodes. This approach may negatively affect availability and data integrity, but also improve privacy~\cite{xu2017taxonomy}. Corda is not the only blockchain project addressing particular requirements. Since the advent of Bitcoin in 2008, a diverse range of blockchains has emerged, resulting in solutions  with many different features and configurations. Differences range from cost efficiency, storage and performance to decentralization and access restrictions~\cite{xu2017taxonomy}.

Due to their various features and properties, different blockchains are not equally suitable for a given use case, inevitably leading to the question which blockchain meets the requirements of a user to the largest extent~\cite{xu2017taxonomy}. As described above, blockchain properties vary over time. Such variations may impact the suitability of a particular blockchain for a given use case. Assuming a decentralized application relied on Bitcoin Cash during the ``hash war'', application users would have been completely reliant on a few miners, contradicting the requirement of decentralization. In such a case, it may be appropriate to switch to another blockchain providing similar features along with a higher degree of decentralization. Furthermore, also user requirements may change over time and thus, another blockchain may become more appropriate for a certain use case~\cite{changing-user-requirements}, e.g., demanding access restrictions offered by permissioned blockchains.

Summarizing, engineers seeking to utilize a blockchain for their applications face a diverse range of blockchain technologies with different features and configurations. Furthermore, a former technology decision may become outdated due to variations of blockchain properties or changing user demands. In order to overcome these issues, a solution is required that monitors multiple blockchains, determines the most appropriate one based on user preferences, and enables a switchover between blockchains at runtime. Such a framework is supposed to continuously monitor several blockchains. In case another blockchain becomes more appropriate than the currently used one, the framework is expected to suggest switching to that chain, i.e., to route subsequent operations (e.g., reading or writing data) to the new blockchain. Additionally, during the switchover, a user-defined amount of data stored on the currently used blockchain could be moved to the target chain. This, for instance, may be essential if a certain amount of data is needed on the target blockchain for further processing or in case the community is losing trust in the currently used blockchain. In the next section, we will introduce how our framework implements the needed functionalities.

%% file: sections/03_approach.tex
\section{Approach}
\label{sec:approach}
In order to address the requirements outlined in the previous section, the proposed framework consists of three main components: The \textit{Monitoring Component} continuously surveys information about each supported blockchain and calculates metric values. Based on these metric values, the \textit{Blockchain Selection Algorithm} calculates each blockchain's benefit and selects the most beneficial one. In case another blockchain is more beneficial than the currently used one, a switchover is suggested. The \emph{Switchover Component} provides the possibility to switch from one blockchain to another.

In the following, we first introduce in Section~\ref{sec:metrics} the blockchain metrics supported by the Monitoring Component. In Section~\ref{sub:blockchainselection}, we discuss the Blockchain Selection Algorithm, while Section~\ref{sec:switchover} presents the functionality of the Switchover Component. Finally, the technical design of the framework is discussed in Section~\ref{sub:technicaldesign}. 

\subsection{Blockchain Metrics}
\label{sec:metrics}
In order to select the most appropriate blockchain, users should be able to define particular selection metrics, which are then applied in order to assess how a blockchain matches the needs of the user. In the following, we present eight blockchain metrics relevant for the comparison of different blockchains. They can be categorized into cost-related metrics (M1-3), performance-related metrics (M4-5), security-related metrics (M6-7), and reputation (M8). Notably, the framework presented in this paper allows to extend the metric model by further metrics, if necessary. Therefore, the discussed metrics should be considered as exemplary. The main requirement for a metric to be added to the metric model is that it is measurable (which is the case for the cost, performance, and security metrics below) or can be defined by the user (which is the case for reputation metrics).

\subsubsection{Cost of writing 1 KB of data into a blockchain~(M1)}
This metric represents the cost of writing one kilobyte of data into a blockchain (in USD). Since many blockchains allow their users to prioritize transactions by specifying transaction fees, the cost may vary depending on the fees the user is willing to pay. The calculation is based on the transaction fees provided by the framework user. In case no fees are provided, the framework automatically determines fees that will cause submitted transactions to get included within a predefined number of blocks. Since we exemplarily use Bitcoin, Ethereum, Ethereum Classic, and Expanse in the reference implementation, we set this number to six as a trade-off between cost and performance. In case other blockchains are connected to the framework, a different block number may be more appropriate. By using the introduced metric, the framework can determine the cheapest blockchain. We have selected one kilobyte as basis for the calculation of this metric, since this amount is sufficient to store various kinds of meta data, e.g., log events. In case another amount is more appropriate, it can be changed without any restrictions.

\subsubsection{Cost of retrieving 1 KB of data from a blockchain~(M2)}
This metric specifies the cost of retrieving one kilobyte of data from a blockchain (in USD). Typically, reading data from a blockchain is free of charge. Nevertheless, we introduce this metric since the proposed framework is intended to be applicable for a wide range of use cases, including possible invocations of non-read-only smart contract methods for retrieving data from a blockchain (e.g., access logging). This metric enables the framework to compare different blockchains regarding cost of retrieving data. Analogous to M1, the amount of data used for the calculation can be changed without any restrictions.

\subsubsection{Exchange rates (M3)}
This metric represents the current exchange rate between USD and the native cryptocurrency of a particular blockchain, e.g., the market price for one Bitcoin in USD. Exchange rates are required for calculating the cost of interacting with a particular blockchain.

\subsubsection{Inter-block time (M4)}
The inter-block time specifies the rolling average of the time (in seconds) it takes to mine a block and is calculated on the basis of all blocks that have been mined during the last 24 hours. The inter-block time is used as an indicator of a blockchain's performance.

\subsubsection{Transaction throughput (M5)}
The transaction throughput represents the rolling average of the number of transactions that are processed per second and is calculated based on all transactions that have been mined during the last 24 hours. Analogous to M4, this metric is used to observe a blockchain's performance.

\subsubsection{Degree of decentralization (M6)}
This metric specifies the distribution of the network's hash power among miners or mining pools. The framework provides a mapping between miner addresses and their proportion of mined blocks. The proportion is specified in percent and is calculated from the blocks that have been mined during the last 24 hours. In case of Ethereum-based blockchains, uncle blocks are also taken into account. This metric allows the identification of miners that control large amounts of hash power. Miners controlling more than 50\% of a network's hash power can tamper with the blockchain, since they are able to generate more blocks, enabling them to master the longest chain~\cite{blockchain-consensus-algorithm}.

\subsubsection{Network hash rate (M7)}
This metric specifies the hash rate the network has performed in the recent 24 hours. The hash rate is computed from the current difficulty and from the blocks that have been mined during the last 24 hours. In case of Ethereum-based blockchains, uncle blocks are also taken into account. This metric allows to observe the progression of the network's hash rate, enabling the identification of significant declines.

\subsubsection{Reputation (M8)}
The reputation is an integer value between 0 and 10, and indicates the degree of renown a blockchain is associated with. It may reflect various properties such as trust, frequency of new feature releases, number of forks, community consensus and controversies, security concerns, etc. The value 0 indicates the worst reputation, whereas the value 10 represents an excellent reputation. This metric is introduced to compare different blockchains by their renown.

Metrics M1--M7 are calculated automatically by the framework's Monitoring Component. For M8, manual user input is required, since this metric highly depends on the subjective assessment of the framework user.

An overview of each metric's data type is given in Table \ref{tab:metric-datatypes}.

\begin{table}
	\centering
	\caption{Metric data types.}
	\begin{tabular}{c l}
		\toprule
		\textbf{Metric} & \textbf{Data type} \\
		\midrule
		M1 & Decimal $\geq 0$ \\
		M2 & Decimal $\geq 0$ \\
		M3 & Decimal $\geq 0$ \\
		M4 & Decimal $\geq 0$ \\
		M5 & Decimal $\geq 0$ \\
		M6 & Mapping (key: string, value: decimal $\geq 0$) \\
		M7 & Decimal $\geq 0$ \\
		M8 & Integer $\geq 0$ and $\leq 10$ \\
		\bottomrule
	\end{tabular}
	\label{tab:metric-datatypes} 
\end{table}

\subsection{Blockchain Selection Algorithm}
\label{sub:blockchainselection}
Next, we introduce concepts for comparing different blockchains and for selecting the most appropriate one. To give the user the opportunity to define which metrics are of high and low importance, respectively, we first introduce a weighted ranking system used to calculate a blockchain's benefit. For this, each blockchain metric is assigned a user-defined weight indicating its importance when calculating an overall ranking of blockchains. Table~\ref{tab:weights-meaning} shows six possible weights offered by the framework.

\begin{table}
	\centering
	\caption{Weights used by the framework.}
	\begin{tabular}{c l}
		\toprule
		\textbf{Weight} & \textbf{Meaning} \\
		\midrule
		0 & No importance\\
		1 & Very low importance\\
		2 & Low importance\\
		3 & Medium importance\\
		4 & High importance\\
		5 & Very high importance\\
		\bottomrule
	\end{tabular}
	\label{tab:weights-meaning} 
\end{table}

Furthermore, the user has to specify a \emph{Score Assignment Function} (SAF) for each blockchain metric. The SAF maps a concrete metric value to a score, as shown in (\ref{eq:score-assign-function}). $D_i$ denotes the data type of metric $i$.

\begin{equation}\label{eq:score-assign-function}
\text{SAF}: D_i \mapsto \{0,1,2,3,4\}
\end{equation}

The SAF is applied in order to normalize different data types and ranges of data. Through this, it is possible to combine the different metrics and use them for a weighted ranking of blockchains, despite the fact that the data ranges and data types of the single metrics differ. 

The score calculated by the SAF quantifies how well a metric satisfies a certain property, e.g., an inter-block time of 100 seconds may be rewarded with a score of 2. A complete example for how a SAF could be defined by the user is given in Section~\ref{sec:eval}. 

In Table~\ref{tab:metric-scores}, five possible score values and their meanings are presented. The SAF of a metric is applied to the corresponding metric value of each supported blockchain. Assuming the framework supports Bitcoin and Ethereum, the SAF of M1 is applied to the cost of writing data into the Bitcoin blockchain and to the cost of writing data into the Ethereum blockchain. In a next step, the SAF of M2 is applied to the cost of reading data from the Bitcoin blockchain and to the cost of reading data from the Ethereum blockchain. The same procedure is repeated for  M3--M8. In the presented example, the entire process results in two score values for each metric (one for Bitcoin and one for Ethereum). By providing weight and score assignments, the user is able to customize the internal logic of the framework to meet desired needs.

\begin{table}
	\centering
	\caption{Score definitions used by the framework.}
	\begin{tabular}{c l}
		\toprule
		\textbf{Score} & \textbf{Meaning} \\
		\midrule
		0 & Does not satisfy \\
		1 & Partly satisfies \\
		2 & Substantially satisfies \\
		3 & Almost satisfies \\
		4 & Fully satisfies \\
		\bottomrule
	\end{tabular}
	\label{tab:metric-scores} 
\end{table}

The benefit of a blockchain $\mathcal{B}$ is calculated by summing up each metric's weighted score, as shown in (\ref{eq:blockchain-benefit}), where $n$ denotes the number of blockchain metrics, $\text{weight}_i$ represents the user-defined weight of metric $M_i$ and $\text{score}_{\mathcal{B}[i]}$ dubs the score value of metric $M_i$ for blockchain $\mathcal{B}$ (obtained by applying the SAF to the value of metric~$M_i$ for $\mathcal{B}$).

\begin{equation}\label{eq:blockchain-benefit}
\sum_{i=1}^{n}\text{weight}_i \cdot \text{score}_{\mathcal{B}[i]}
\end{equation}

The presented formula is applied for each supported blockchain. The blockchain with the highest benefit is chosen as the most beneficial one. Table \ref{tab:example-weighted-ranking} shows an example of a weighted ranking, where two blockchains are evaluated. Blockchain A has a benefit (total weighted score) of 103, whereas Blockchain B has a benefit of 101. Therefore, blockchain A is considered as more beneficial.

\begin{table}
	\centering
	\caption{An example of a weighted ranking with two blockchains.}
	\renewcommand{\arraystretch}{1.2}
	\begin{tabular}{cc|c|c|c|c|c|c|}
		\cline{3-6}
		& & \multicolumn{2}{ c| }{\textbf{Blockchain A}} & \multicolumn{2}{ c| }{\textbf{Blockchain B}} \\ \cline{1-6}
		\multicolumn{1}{ |c }{\textbf{Metric}} & \multicolumn{1}{ |c| }{\textbf{Weight}} & \textbf{Score} & \textbf{W. Score} & \textbf{Score} & \textbf{W. Score} \\ \cline{1-6}
		\multicolumn{1}{ |c  }{M1} & \multicolumn{1}{ |c| }{5} & 4 & 20 & 3 & 15 \\ \cline{1-6}
		\multicolumn{1}{ |c  }{M2} & \multicolumn{1}{ |c| }{3} & 4 & 12 & 4 & 12 \\ \cline{1-6}
		\multicolumn{1}{ |c  }{M3} & \multicolumn{1}{ |c| }{4} & 4 & 16 & 2 & 8 \\ \cline{1-6}
		\multicolumn{1}{ |c  }{M4} & \multicolumn{1}{ |c| }{5} & 2 & 10 & 4 & 20 \\ \cline{1-6}
		\multicolumn{1}{ |c  }{M5} & \multicolumn{1}{ |c| }{3} & 3 & 9 & 3 & 9 \\ \cline{1-6}
		\multicolumn{1}{ |c  }{M6} & \multicolumn{1}{ |c| }{3} & 3 & 9 & 3 & 9 \\ \cline{1-6}
		\multicolumn{1}{ |c  }{M7} & \multicolumn{1}{ |c| }{5} & 3 & 15 & 4 & 20 \\ \cline{1-6}
		\multicolumn{1}{ |c  }{M8} & \multicolumn{1}{ |c| }{4} & 3 & 12 & 2 & 8 \\ \cline{1-6}
		\multicolumn{1}{ |c  }{\textbf{Total}} & \multicolumn{1}{ |c| }{\textbf{32}} & \textbf{26} & \textbf{103} & \textbf{25} & \textbf{101} \\ \cline{1-6}
	\end{tabular}
	\label{tab:example-weighted-ranking} 
	\renewcommand{\arraystretch}{1}
\end{table}

The introduced weighted ranking system allows the quantification of a blockchain's benefit on the basis of user-defined weights and score assignments. However, it does not offer a mechanism for enforcing certain requirements (e.g., an inter-block time lower than or equal to 60 seconds) a blockchain \emph{must} fulfill under all circumstances, regardless of its benefit. The example shown in Table \ref{tab:example-weighted-ranking} outlines a situation where Blockchain A has a worse score for M4 (inter-block time) than Blockchain B, but A is still the most beneficial one due to the scores of other metrics. 

However, it might be the case that it is of utmost importance to the user that for a particular metric a certain threshold is met, e.g., in the example just mentioned, that the inter-block time (M4) has at most a value of 60 seconds. Thus, the Blockchain Selection Algorithm is adapted to consider only those blockchains that fulfill such additional requirements, i.e., only those blockchains that satisfy these additional requirements serve as candidates for the weighted ranking system. All other blockchains are not further regarded.

For that purpose, we introduce the \emph{Metric Validation Function} (MVF). As shown in (\ref{eq:metric-validation-function}), this function maps an 8-tuple to a 9-tuple. $D_i$ represents the data type of metric $M_i$.
\begin{equation}\label{eq:metric-validation-function}
\text{MVF}: D_1 \times \dots \times D_8 \mapsto \{\textit{true}, \textit{false}\}^9
\end{equation}
The left-most~(first) value of the returned 9-tuple indicates whether metric M1 is valid (i.e., whether it satisfies user requirements), the second value indicates whether metric M2 is valid and so forth. The last~(right-most) value of the 9-tuple represents the validity of a blockchain and may be the result of a combination of the single boolean values, e.g., linking the single boolean values to a propositional formula. This enables the user to specify more complex requirements. The single boolean values are needed for further decisions during the switchover (see Section \ref{sec:switchover}). The concrete implementation of the proposed function has to be provided by the user.

\subsection{Switchover Functionality}
\label{sec:switchover}

As described in Section~\ref{sec:background}, a switchover is the process of routing all subsequent operations (e.g., read and write operations) to another blockchain. Depending on user preferences, the switchover is either performed fully automated once a more beneficial blockchain is detected, or has first to be approved by the user.

Furthermore, the framework allows the user to define the amount of already existing data records that should be moved from the currently used blockchain to the destination chain during a switchover. This amount may depend on the metric(s) that caused the switchover. For instance, if the community is losing trust in the currently used blockchain, it may be essential to transfer \emph{all} data stored on the currently used blockchain or at least data of a specific period of time to the destination blockchain. 

In order to customize the framework's logic for determining the amount of data that should be transferred, the user can specify a custom strategy. Whenever a more appropriate blockchain is detected, i.e., a switchover is suggested by the framework, this custom strategy is triggered. The framework forwards each metric's weighted score and the validation results (obtained from the MVF) of both the currently used blockchain and the suggested chain to the user-defined strategy. The presence of this information enables the user to define a strategy that is able to determine the amount of data on the basis of those metrics that cause the currently used chain to be less appropriate than the suggested one. 

The amount of data to be transferred is specified by a date range. For instance, if the user-defined strategy specifies a range between 01.02.2019 and 28.02.2019, all data records mined during this range are copied to the destination chain. It should be noted, that -- while we use the term ``transfer data'' to describe that data is copied from one blockchain to another -- data can of course not be deleted from the original blockchain. The goal of data transfers is merely to make sure that no data gets lost. For instance, if the number of miners for a particular blockchain rapidly decreases and this is reflected in a very low degree of decentralization (M6), the chance is very high that a malicious, powerful attacker can perform a 51\% attack on the blockchain, thus rendering the data in the blockchain useless.

In order to prevent the framework from performing multiple switchovers within a short period of time, e.g., due to frequent variations of the order in the weighted ranking system, we introduce a \textit{switchover suppression period}. This period can be defined by the user. The framework suppresses subsequent switchover suggestions until the switchover suppression period elapses, regardless of how many changes are occurring in the ranking system. Thus, at most one switchover is suggested every period, preventing immediate switchovers between blockchains. If no timespan is defined, the framework starts the switchover immediately after a more beneficial blockchain has been detected (and automatic switchovers are enabled).

\subsection{Technical Design}
\label{sub:technicaldesign}

\begin{figure*}[t]
	\centering
	\includegraphics[width=\textwidth]{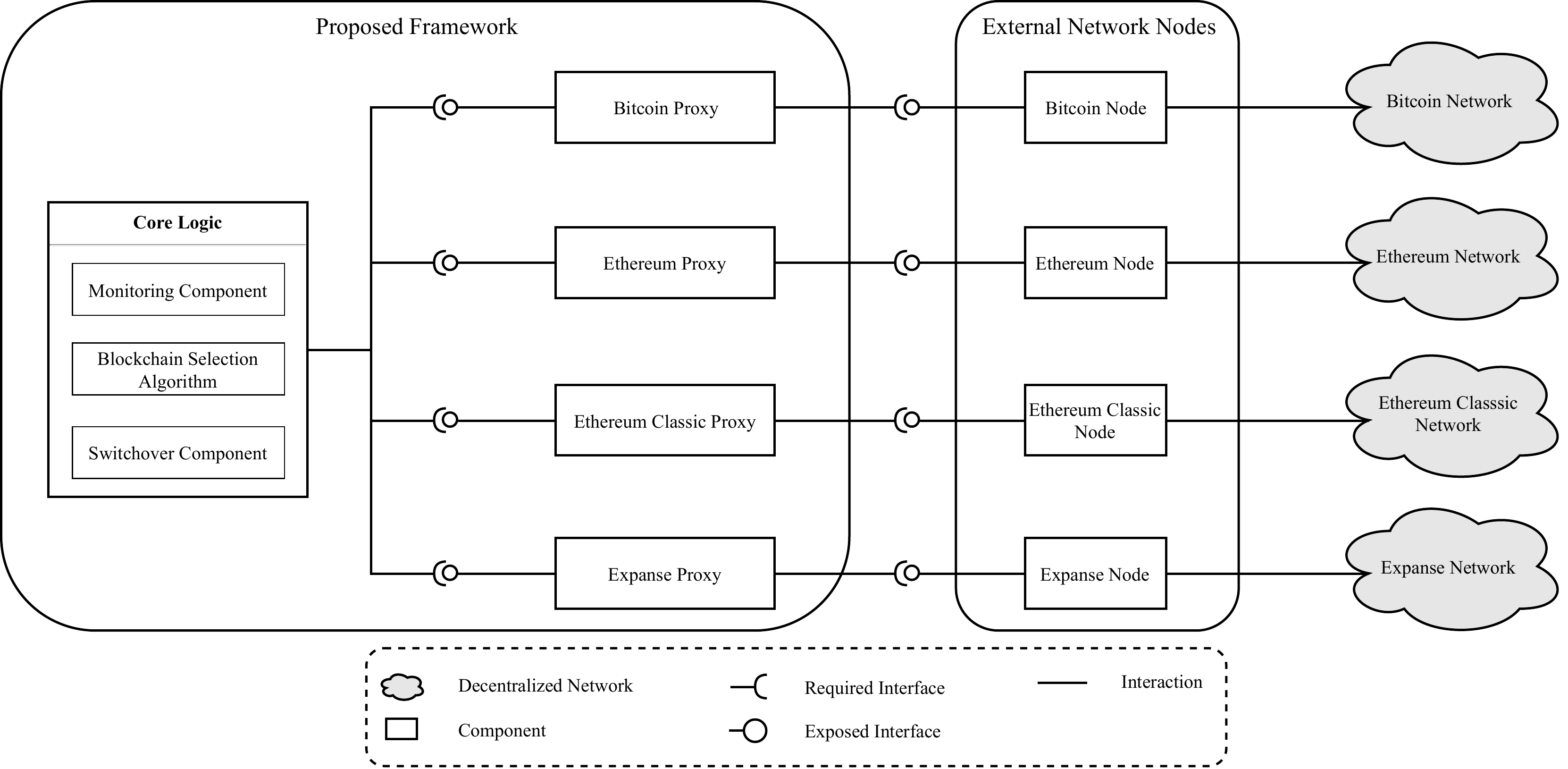}
	\caption{Architecture of the proposed framework.}
	\label{fig:framework-architecture-overview}
\end{figure*}

Figure~\ref{fig:framework-architecture-overview} presents an overview of the framework's architecture. The proposed framework consists of the \emph{Core Logic} and a number of \emph{Blockchain Proxies}. The depicted external network nodes serve as bridges between the framework and blockchain networks (e.g., the Ethereum network). They are used by the framework for interacting with a blockchain's network, e.g., for submitting new transactions or requesting new blocks.

The Core Logic communicates with Blockchain Proxies and consists of the three major components presented in Sections~\ref{sec:metrics} to \ref{sec:switchover}: The Monitoring Component, the Blockchain Selection Algorithm, and the Switchover Component. Based on data requested via the Blockchain Proxies, the Core Logic validates each blockchain's metrics by applying MVF, calculates each blockchain's benefit, determines the most beneficial chain, and provides the functionality for switching to another blockchain. The Core Logic is agnostic to a blockchain's technical details. Instead, the according Blockchain Proxy translates data from a particular blockchain into a neutral format that can be processed by the Core Logic. For each supported blockchain, such a proxy is implemented. A proxy abstracts interactions with the underlying blockchain by providing an interface used by the Core Logic. Examples for interactions are writing new data records into a blockchain, reading data records from a blockchain, and requesting new blocks. 

We have selected Bitcoin, Ethereum, Ethereum Classic, and Expanse for the prototypical implementation of the proposed framework\footnote{\url{https://github.com/pf92/blockchain-interop}}. The first three blockchains have been chosen due to their popularity, whereas Expanse has been selected since storing data is very cheap due to the low market price. Notably, the proposed framework is not restricted to these blockchains. The mentioned blockchains should be considered as exemplary. In case further blockchains should be supported, the framework can be extended by providing additional proxy implementations.

In order to save disk space, the framework only keeps blocks in memory that are needed for calculating each blockchain's metrics, i.e., only blocks which have been mined during the last 24 hours are stored.

The framework's design incorporates the reactive programming paradigm. In the reactive programming paradigm, data flows and the propagation of changes play a key role. If a data source changes its value, the change is propagated through the entire topology, i.e., each operator or observer that is part of the topology or is registered to receive notifications is informed about changes \cite{reactive-programming}. In the framework's architecture, external network nodes act as data sources. Once a new block is received by the framework, subsequent computation steps are triggered in order to recalculate the metrics discussed in Section~\ref{sec:metrics}. Each new block affects the calculation of the metric values, and changes of metric values affect the blockchain selection algorithm, i.e., the weighted ranking system and the MVF, as discussed in Section~\ref{sub:blockchainselection}.

In the following, we further elaborate on the calculation of each metric, taking into account the four blockchains which we have selected for the prototypical implementation of our framework, i.e., Bitcoin, Ethereum, Ethereum Classic, and Expanse.

\subsubsection{Cost of writing 1 KB of data into a blockchain~(M1)}
A cheap method for storing data on the Bitcoin blockchain is to use the script operation code OP\_RETURN~\cite{data-insertion-bitcoin}. OP\_RETURN marks a transaction output as invalid and accepts a user-defined sequence of up to $80$ bytes~\cite{btc-op-return,data-insertion-bitcoin}. In order to write data records into the Bitcoin blockchain, a transaction with one input and two outputs (one output that spends the remaining coins and another one that holds the data) is sufficient. A transaction that stores $80$ bytes of data in its second output has an overall size of $282$ bytes. To store one kilobyte of data, $13$ transactions are required. The overall size of these $13$ transactions is 12 $\cdot$ 282 + 266 = 3,650. $266$ bytes is the size of the transaction that stores the remaining $64$ bytes. The overall size of 3,650 bytes is multiplied with the user-defined transaction fees (Satoshi per byte). If the user does not provide transaction fees, an estimation of transaction fees is requested from external APIs.

Ethereum-based blockchains like the selected Ethereum, Ethereum Classic, and Expanse, offer three possibilities for storing data on the blockchain~\cite{xu2017taxonomy}. The first possibility is to store records in the data field of a transaction. In Ethereum and Ethereum Classic (both systems use the Ethereum Virtual Machine), every transaction costs 21,000 gas. For every non-zero byte that is stored in a transaction's data field, additional 68 gas have to be paid~\cite{ethereum-yellow-paper}. The number of bytes stored in a transaction is bounded by the current block gas limit. A single transaction carrying one kilobyte of data would consume at most 21,000 + 68 $\cdot$ 1024 = 90,632 gas. As of February 2019, the block gas limit for Ethereum and Ethereum Classic is about 8,000,000, far enough for executing transactions holding one~kilobyte of data. 

The second possibility is to store data in logs. The cost of storing one byte in a log is eight gas. Additional $375$ gas have to be paid for the LOG operation~\cite{ethereum-yellow-paper}. However, this option requires a smart contract that emits log events. Due to different ways of writing such a contract, it is difficult to calculate the cost. Nevertheless, the second option is more expensive than the first, since every input data intended to be logged as event is also encoded in the data field of the transaction submitted for the contract call. 

The third option is to store data in a smart contract's storage. Storing a 32-byte word in a smart contract's storage costs 20,000 gas~\cite{ethereum-yellow-paper}. Additional gas has to be paid for the transaction that contains the contract call and the input data. 

Therefore, the first option is the cheapest one. This statement also holds for Expanse, since the aforementioned operations consume the same gas cost on the Expanse Virtual Machine\footnote{\url{https://expanse.tech/docs/developer/}}. In case the user does not specify a preferred gas price, the framework requests the median gas price from the external network nodes.

\subsubsection{Cost of retrieving 1 KB of data from a blockchain~(M2)}
Since data records are stored in a transaction's data field without the involvement of smart contracts, reading data records is free of charge. Hence, in the current reference implementation, this metric is always zero for all supported blockchains.

\subsubsection{Exchange rates (M3)}
The framework continuously requests the current market price in USD for cryptocurrencies associated with the supported blockchains. In the reference implementation, we use CryptoCompare\footnote{\url{https://min-api.cryptocompare.com/}}, an external service exposing interfaces for requesting these market prices.

\subsubsection{Inter-block time (M4)}
The rolling average of the time between two blocks is computed by applying the formula shown in (\ref{eq:block-time}), where \textit{n} denotes the number of blocks mined during the last 24 hours. The presented formula is applied for all supported blockchains.
\begin{equation}\label{eq:block-time}
\dfrac{24 \cdot 3600}{n}
\end{equation}

\subsubsection{Transaction throughput (M5)}
As shown in (\ref{eq:tx-throughput}), the transaction throughput (rolling average) is computed by summing up the number of transactions stored in each block that has been mined during the last 24 hours, and by dividing this sum by $24 \cdot 3600$. \textit{n} denotes the total number of mined blocks during the last 24 hours and $\text{txcount}_i$ represents the number of transactions of block $i$.
\begin{equation}\label{eq:tx-throughput}
\dfrac{\sum_{i=1}^{n}\text{txcount}_i}{24 \cdot 3600}
\end{equation}

\subsubsection{Degree of decentralization (M6)}
We calculate the distribution of a network's hash power in two different ways due to fundamental differences between Bitcoin- and Ethereum-based blockchains. In Bitcoin, it is sufficient to count the number of blocks for each miner that has mined at least one block during the last 24 hours and, in a further step, to divide each miner's block counter by the overall number of blocks. For Ethereum-based blockchains, in addition to regular blocks, also uncle blocks are taken into account, since miners also spend computational power for integrating these blocks.

\subsubsection{Network hash rate (M7)}
For Bitcoin, the average number of hashes per second the network has performed in the last 24 hours is computed as shown in (\ref{eq:btc-network-hashrate}).
\begin{equation}\label{eq:btc-network-hashrate}
\dfrac{n}{144} \cdot \dfrac{D \cdot 2^{32}}{600}
\end{equation}
Here, $n$ denotes the number of blocks that have been mined during the last 24 hours. $D \cdot 2^{32}$ specifies the expected number of hashes that have to be calculated to find a block with difficulty $D$~\cite{btc-difficulty}. In Bitcoin, $D$ is set such that, on average, a new block is mined every ten minutes (600 seconds)~\cite{btc-difficulty}. Thus, $144$ is the number of blocks are expected to get mined within 24 hours. Since a new block is anticipated to get mined every ten minutes, $D \cdot 2^{32}$~hashes are expected to be computed in 600~seconds, yielding an average network hash rate of $\dfrac{D \cdot 2^{32}}{600}$~hashes per second~\cite{btc-difficulty}. The term $\dfrac{n}{144}$ adjusts this hash rate in case less than or more than $144$ blocks have been mined. We calculate the hash rate of Ethereum-based networks by summing up the \emph{difficulty} field of each block and each uncle block mined during the last 24 hours and by dividing this sum by the number of seconds equal to 24 hours.

%% file: sections/04_eval.tex
\section{Evaluation}
\label{sec:eval}
In order to evaluate the proposed framework, we investigate its benefit in terms of cost, performance, and trust by using exemplary scenarios. For this, we analyze the framework's reaction to varying blockchain properties as well as its handling of changing user demands.

Since the framework relies on external network nodes, we set up a Bitcore\footnote{\url{https://bitcore.io/}} node for Bitcoin, two Parity\footnote{\url{https://www.parity.io/ethereum/}} nodes for Ethereum and Ethereum Classic, and we use gexp\footnote{\url{http://expanse-org.github.io/go-expanse/}} to run an Expanse network node. Each deployed network node uses the respective main chain. For our experiments, we deploy each network node on a separate virtual machine (1 vCPU, 4 GB~RAM) hosted on the Google Cloud Platform. The framework itself is executed on a MacBook Pro (late 2013, 2.4~GHz Intel Core i5, 8 GB 1600 MHz DDR3, Intel Iris 1536 MB, 256 GB~SSD, macOS 10.13.6, Oracle JDK 10).

In the following, four evaluation scenarios are presented. Scenario 1 analyzes exemplarily the framework's reaction to varying blockchain metrics by emulating a decreasing hash rate. For this experiment, we select Expanse as the currently used blockchain. The experiment can also be performed with other blockchains such as Bitcoin, Ethereum, or Ethereum Classic without any restrictions. To customize the framework's internal logic, we provide an implementation of the MVF that returns the 9-tuple (true, true, true, true, true, true, false, true, false) in case the network hash rate drops below 180 GH/s, otherwise (true, true, true, true, true, true, true, true, true). Thus, if the network hash rate drops below 180 GH/s, M7 and the corresponding blockchain are invalid (denoted by the boolean value \emph{false}). Furthermore, we set each metric's weight to 1 and provide for each metric a score assignment function that always returns a score value of 1, since for the conduction of this experiment it is not relevant which blockchain is selected by the framework after the detection of a decreasing hash rate. Starting at a hash rate of 200~GH/s, we emulate a decreasing hash rate reduced by 5~GH/s every 5 seconds. As shown in Listing \ref{lst:log-dropping-hashrate}, once the hash rate is under 180~GH/s, the framework suggests to switch to another blockchain (here:~Ethereum Classic).

\begin{lstlisting}[float, basicstyle=\fontsize{7}{10}\selectfont\ttfamily, caption=Log extraction that shows the reaction of the framework in case the network hash rate of Expanse decreases rapidly., label=lst:log-dropping-hashrate, escapechar=|]
13:52:25,189 - Switchover suggestion: Expanse
13:52:26,983 - Expanse network hash rate: 195.0 GH/s
13:52:31,984 - Expanse network hash rate: 190.0 GH/s
13:52:37,806 - Expanse network hash rate: 185.0 GH/s
13:52:42,807 - Expanse network hash rate: 180.0 GH/s
13:52:46,844 - Expanse network hash rate: 175.0 GH/s
13:52:46,845 - Hash rate (175.0 GH/s) violated 
13:52:46,847 - Switchover suggestion: Ethereum Classic 
\end{lstlisting}

For Scenarios 2--4, we assume that the framework is used in a service-oriented architecture that is made up of different services adopted and operated by several independent and possibly competing business partners. In order to monitor the adherence to service-level agreements (SLAs), services publish relevant information to a blockchain. We conduct these scenarios based on metric values measured between 25.09.2018 and 17.10.2018.

Scenario 2 analyzes the benefit of the framework's selection mechanism in terms of cost, performance, and trust. We assume the involved business partners want to use a blockchain that is cheap, fast and has a high level of trust. The framework is configured with the weighted ranking settings outlined in Table~\ref{tab:eval-ranking}. Since we assume a demand for very cheap and fast write operations, and a high level of trust, we set the weights for M1, M3, M4, M5, M6, M7 and M8 to five (highest importance). Metric M2 can be ignored (i.e., we set the corresponding weight to zero), since read operations are free and our reference implementation does not make use of smart contracts. In order to benefit from an accurate selection, we define the score assignments as granularly as possible, e.g., by considering very low cost in the score assignment.

\begin{table}
	\centering
	\caption{Weighted ranking settings used for the evaluation (M2 is not listed since it is always zero in the implementation).}
	\begin{tabular}{l l l l}
		\toprule
		\textbf{Metric} & \textbf{Weight} & \textbf{Score Assignment} \\
		\midrule
		M1 & 5 & \makecell[l]{$[0; 10^{-4}) \rightarrow 4$, $[10^{-4}; 10^{-2}) \rightarrow 3$, \\ $[10^{-2}; 10^{-1}) \rightarrow 2$, $[10^{-1}; 1) \rightarrow 1$, \\ $[1; \infty) \rightarrow 0$ \\} \\
		\midrule
		M3 & 5 & \makecell[l]{$[0; 50) \rightarrow 4$, $[50; 100) \rightarrow 3$, \\ $[100; 250) \rightarrow 2$, $[250; 500) \rightarrow 1$, \\ $[500; \infty) \rightarrow 0$ \\} \\
		\midrule
		M4 & 5 & \makecell[l]{$[0; 20) \rightarrow 4$, $[20; 40) \rightarrow 3$, \\ $[40; 60) \rightarrow 2$, $[60; 120) \rightarrow 1$, \\ $[120; \infty) \rightarrow 0$ \\} \\
		\midrule
		M5 & 5 & \makecell[l]{$[10; \infty) \rightarrow 4$, $[5; 10) \rightarrow 3$, \\ $[2; 5) \rightarrow 2$, $[0.45; 2) \rightarrow 1$, \\ $[0; 0.45) \rightarrow 0$ \\} \\
		\midrule
		M6 &5 & \makecell[l]{Proportion (\%) of the biggest miner: \\ $[0; 22) \rightarrow 4$, $[22; 27) \rightarrow 3$, \\ $[27; 30) \rightarrow 2$, $[30; 38) \rightarrow 1$, \\ $[38; \infty) \rightarrow 0$ \\} \\
		\midrule
		M7 & 5 & \makecell[l]{Rates are denoted in terahashes: \\ $[1,000; \infty) \rightarrow 4$, $[700; 1,000) \rightarrow 3$, \\ $[400; 700) \rightarrow 2$, $[100; 400) \rightarrow 1$, \\ $[0; 100) \rightarrow 0$ \\} \\
		\midrule
		M8 & 5 & \makecell[l]{$[8; 10] \rightarrow 4$, $[6; 8) \rightarrow 3$, \\ $[4; 6) \rightarrow 2$, $[2; 4) \rightarrow 1$, $[0; 2) \rightarrow 0$ \\} \\
		\bottomrule
	\end{tabular}
	\label{tab:eval-ranking} %
\end{table}

According to their popularity and miner activity, we assume a reputation of 10 for Bitcoin and Ethereum, a reputation of 9 for Ethereum Classic and a reputation of 5 for Expanse. In order to emulate an execution on 25.09.2018, the framework is fed with the values measured on that day. According to the weighted ranking outlined in Table~\ref{tab:eval-weighted-ranking-1}~(metrics with a weight of zero are omitted), Ethereum is the most beneficial blockchain. The key points of this selection are as follows: 
\begin{itemize}
	\item By using Ethereum, the cost of writing one KB of data is approximately 24 times lower than the cost of writing the same amount of data into the Bitcoin blockchain. The cost of writing data into the Ethereum Classic and Expanse blockchains is about 154 times and about 96 times lower, respectively, than the cost when using Ethereum.
	\item Compared to Bitcoin with a price of 6,394.25 USD, the exchange rate of Ethereum is about 30 times lower. The price in USD for one Ether is approximately 20 times greater than the price for one token on Ethereum Classic. With an exchange rate of 0.36 USD, Expanse is the cheapest token.
	\item Ethereum features an inter-block time about 38 times shorter than Bitcoin and about three times shorter than Expanse. Ethereum Classic has almost the same inter-block time as Ethereum.
	\item With a throughput of about 5.74 transactions per second~(tps), Ethereum processes by far the greatest number of transactions, whereas Bitcoin has a rate of 2.57 tps, Ethereum Classic a rate of 0.47 tps, and Expanse handles only 0.06 tps.
	\item The network hash rate of Ethereum is about 16 times greater than the rate of Ethereum Classic and about 1,275 times greater than the hash rate of Expanse. The Bitcoin network mines with a hash rate approximately 217,012 times greater than that of Ethereum.
	\item The biggest Ethereum miner controls about 24\% of the network's hash power, whereas the biggest Ethereum Classic and Expanse miners control about 42\% and 48\%, respectively. The biggest miner of the Bitcoin network controls about 20\%.
\end{itemize}

\begin{table}
	\centering
	\caption{Weighted ranking of Scenario 2 (based on values measured on 25.09.2018).}
	\renewcommand{\arraystretch}{1.2}
	\begin{tabular}{c|c|c|c|c|}
		\cline{1-5}
		\multicolumn{1}{ |c| }{\textbf{Metric}} & \makecell[c]{\textbf{Bitcoin} \\ \textbf{W. Score}} & \makecell[c]{\textbf{Ethereum} \\ \textbf{W. Score}} & \makecell[c]{\textbf{Ethereum Classic} \\ \textbf{W. Score}} & \makecell[c]{\textbf{Expanse} \\ \textbf{W. Score}} \\ \cline{1-5}
		\multicolumn{1}{ |c|  }{M1} & 0 & 5 & 15 & 15 \\ \cline{1-5}
		\multicolumn{1}{ |c|  }{M3} & 0 & 10 & 20 & 20 \\ \cline{1-5}
		\multicolumn{1}{ |c|  }{M4} & 0 & 20 & 20 & 10 \\ \cline{1-5}
		\multicolumn{1}{ |c|  }{M5} & 10 & 15 & 5 & 0 \\ \cline{1-5}
		\multicolumn{1}{ |c|  }{M6} & 20 & 15 & 0 & 0 \\ \cline{1-5}
		\multicolumn{1}{ |c|  }{M7} & 20 & 5 & 0 & 0 \\ \cline{1-5}
		\multicolumn{1}{ |c|  }{M8} & 20 & 20 & 20 & 10 \\ \cline{1-5}
		\multicolumn{1}{ |c|  }{\textbf{Total}} & \textbf{70} & \textbf{90} & \textbf{80} & \textbf{55} \\ \cline{1-5}
	\end{tabular}
	\label{tab:eval-weighted-ranking-1} 
	\renewcommand{\arraystretch}{1}
\end{table}

Scenario 3 investigates the framework's handling of changing user demands indicated by adjusted metric weights. We assume that engineers of one business partner plan to run data-intensive tests on their adopted services. Since a large amount of data is scheduled to be written into the blockchain, low cost is preferred. Furthermore, we assume that the reputation can be neglected for the test execution. This scenario is conducted based on the weighted ranking settings outlined in Table~\ref{tab:eval-ranking}. In order to incorporate the changed demands in the weighted ranking system, we set the weights for M6, M7, and M8 to 0. Furthermore, we assume that these changes take effect on 07.10.2018. Table~\ref{tab:eval-weighted-ranking-2} shows the weighted ranking based on the changed settings and the metric values gathered on 07.10.2018 (metrics with a weight of zero are omitted). Since Ethereum Classic has the highest benefit, it is selected as the most beneficial chain. Due to the lack of significant variations of blockchain metrics between 25.09.2018 and 07.10.2018, Ethereum has been the most beneficial chain for the entire date range. By switching from Ethereum to Ethereum Classic, the cost of writing one KB of data has been decreased by a factor of 42. Furthermore, Ethereum Classic has almost the same inter-block time as Ethereum (approximately 14 seconds). Since Expanse has an inter-block time of 44 seconds, Ethereum Classic has been preferred, as shown in Table~\ref{tab:eval-weighted-ranking-2}.

\begin{table}
	\centering
	\caption{Weighted ranking of Scenario 3 (based on values measured on 07.10.2018).}
	\renewcommand{\arraystretch}{1.2}
	\begin{tabular}{c|c|c|c|c|}
		\cline{1-5}
		\multicolumn{1}{ |c| }{\textbf{Metric}} & \makecell[c]{\textbf{Bitcoin} \\ \textbf{W. Score}} & \makecell[c]{\textbf{Ethereum} \\ \textbf{W. Score}} & \makecell[c]{\textbf{Ethereum Classic} \\ \textbf{W. Score}} & \makecell[c]{\textbf{Expanse} \\ \textbf{W. Score}} \\ \cline{1-5}
		\multicolumn{1}{ |c|  }{M1} & 0 & 10 & 15 & 20 \\ \cline{1-5}
		\multicolumn{1}{ |c|  }{M3} & 0 & 10 & 20 & 20 \\ \cline{1-5}
		\multicolumn{1}{ |c|  }{M4} & 0 & 20 & 20 & 10 \\ \cline{1-5}
		\multicolumn{1}{ |c|  }{M5} & 10 & 15 & 5 & 0 \\ \cline{1-5}
		\multicolumn{1}{ |c|  }{\textbf{Total}} & \textbf{10} & \textbf{55} & \textbf{60} & \textbf{50} \\ \cline{1-5}
	\end{tabular}
	\label{tab:eval-weighted-ranking-2} 
	\renewcommand{\arraystretch}{1}
\end{table}

The intention of Scenario~4 is to show the effects of changing user requirements  indicated by adjusted score assignments. For this, we assume that an inter-block time between 30 and 60 seconds is completely sufficient for conducting further service tests. Moreover, the transaction throughput becomes less important for further test executions. Due to the large amount of test data that is written to the blockchain, low cost have still high priority. This scenario is conducted based on the weighted ranking settings of Scenario~3. To reflect the changed user requirements in the weighted ranking settings, the score assignment for M4 is changed to: $[0; 60) \rightarrow 4$, $[60; 120) \rightarrow 3$, $[120; 180) \rightarrow 2$, $[180; 240) \rightarrow 1$, $[240; \infty) \rightarrow 0$. Since the transaction throughput has a lower priority, the weight for M5 is set to 3. We further assume that the weighted ranking settings are changed on 17.10.2018. Based on the metric values measured on 17.10.2018 and the changed settings, the framework selects Expanse as the most beneficial chain, since it has the highest score (as shown in Table~\ref{tab:eval-weighted-ranking-3}). Due to the lack of significant variations of blockchain metrics between 07.10.2018 and 17.10.2018, Ethereum Classic has been the most beneficial chain for the entire date range.  A switchover from Ethereum Classic to Expanse enables a cost reduction by a factor of approximately 45.

Summarizing our evaluation, we see that the framework can select the most appropriate blockchain based on user preferences. Furthermore, it is able to react to volatile blockchain properties and it can handle changing user demands. These features allow users to benefit from low cost, better performance, and a higher level of trust.

\begin{table}
	\centering
	\caption{Weighted ranking of Scenario 4 (based on values measured on 17.10.2018).}
	\renewcommand{\arraystretch}{1.2}
	\begin{tabular}{c|c|c|c|c|}
		\cline{1-5}
		\multicolumn{1}{ |c| }{\textbf{Metric}} & \makecell[c]{\textbf{Bitcoin} \\ \textbf{W. Score}} & \makecell[c]{\textbf{Ethereum} \\ \textbf{W. Score}} & \makecell[c]{\textbf{Ethereum Classic} \\ \textbf{W. Score}} & \makecell[c]{\textbf{Expanse} \\ \textbf{W. Score}} \\ \cline{1-5}
		\multicolumn{1}{ |c|  }{M1} & 0 & 10 & 15 & 20 \\ \cline{1-5}
		\multicolumn{1}{ |c|  }{M3} & 0 & 10 & 20 & 20 \\ \cline{1-5}
		\multicolumn{1}{ |c|  }{M4} & 0 & 20 & 20 & 20 \\ \cline{1-5}
		\multicolumn{1}{ |c|  }{M5} & 6 & 9 & 3 & 0 \\ \cline{1-5}
		\multicolumn{1}{ |c|  }{\textbf{Total}} & \textbf{6} & \textbf{49} & \textbf{58} & \textbf{60} \\ \cline{1-5}
	\end{tabular}
	\label{tab:eval-weighted-ranking-3} 
	\renewcommand{\arraystretch}{1}
\end{table}

%% file: sections/06_related.tex
\section{Related Work}
\label{sec:related}
Despite the fact that blockchain technologies have gained much research momentum in recent years, to the best of our knowledge, there are not too many approaches aiming at providing the means to switch between different blockchains. 

One of the earliest contributions in the field of blockchain interoperability is the atomic cross-chain protocol (ACCS) proposed by TierNolan in 2013~\cite{bitcoinforum-atomicswaps}. This protocol enables users of different cryptocurrencies to swap their assets in an atomic fashion. Further contributions focusing on the transfer of assets are The Atomic Swap Technology (TAST)\footnote{\url{https://pantos.io}}, Tesseract~\cite*{tesseract}, BarterDEX\footnote{\url{https://docs.komodoplatform.com/home-whitepaper.html}}, Metronome\footnote{\url{https://www.metronome.io/}}, and the Republic Protocol\footnote{\url{https://renproject.io/}}. Furthermore, sidechains aim to provide interoperability between two blockchains by locking assets on the source chain and creating them on the target blockchain. Transferred assets can only be used on one blockchain at the same time. Cryptographic proofs ensure that assets have been locked on the source chain, before new ones can be created on the target chain~\cite{buterin-chain-interoperability}.

Beyond the trading of assets, Polkadot\footnote{\url{https://polkadot.network/}}, Cosmos\footnote{\url{https://cosmos.network/}}, and Block Collider\footnote{\url{https://www.blockcollider.org/}} aim to integrate blockchains in a more general way, e.g., by enabling communication between smart contracts located on different blockchains.

A further remarkable contribution in the field of blockchain interoperability is BTCRelay\footnote{\url{http://btcrelay.org/}}, a smart contract running on Ethereum that verifies Bitcoin transactions. The contract acts as bridge between the Bitcoin blockchain and Ethereum smart contracts, enabling users to pay with Bitcoin for using Ethereum DAPPs.

To the best of our knowledge, there are no contributions in the field of selection of blockchains during runtime. The discussed approaches do not integrate a mechanism for selecting the most beneficial blockchain based on user-defined requirements. Furthermore, the presented approaches do not provide a functionality for switching back and forth between several blockchains and for migrating already existing data.

%% file: sections/07_conclusion.tex
\section{Conclusion}
\label{sec:conclusion}

In this paper, we have presented a framework capable of switching back and forth between blockchains at runtime. The proposed framework monitors several blockchains, calculates each blockchain's benefits according to user-defined settings, and determines the most beneficial one. Furthermore, the framework is able to react to variations of blockchain metrics and it can handle changing user demands. We have presented the framework's design in detail, using a reference implementation in Java.

Our evaluation shows that switching to another blockchain can save cost and enable users to benefit from better performance and a higher level of trust. The modular design of the framework allows future researchers to add support for further blockchains by providing additional proxy implementations.

As described in Section~\ref{sec:approach}, the framework is able to copy data from the currently used blockchain to the destination chain during a switchover. However, the proposed framework does not enable the migration of smart contracts between blockchains, allowing the automatic deployment of required smart contracts on the destination chain. Such a feature may be relevant if data records are managed by smart contracts rather than stored in a transaction's data field and will therefore be investigated in our future work. Furthermore, the reference implementation stores data records published to Ethereum, Ethereum Classic, and Expanse in a transaction's data field rather than in a smart contract's storage, leading to slower access times. A possible solution we want to investigate is to improve the time needed for searching data by a tracking of transactions that hold data records.